\documentclass{aipproc}

\usepackage{graphicx} 
\usepackage{amssymb}
\usepackage{amsmath}

\newcommand{\ba}{\begin{array}}
\newcommand{\ea}{\end{array}}
\newcommand{\be}{\begin{equation}}
\newcommand{\ee}{\end{equation}}

\layoutstyle{6x9}

\begin{document}

\title{Nearly tri-bimaximal mixing in the $S_3$ flavour symmetry}
\author{A. Mondrag\'on}
{address={Departamento de F\'{\i}sica Te\'orica,\\
Instituto de F\'{\i}sica, Universidad Nacional Aut\'onoma de M\'exico ,\\
Apdo. Postal 20-364,  01000 M{\'e}xico D.F., \ M{\'e}xico.}}
\author{M. Mondrag\'on}
{address={Departamento de F\'{\i}sica Te\'orica,\\
Instituto de F\'{\i}sica, Universidad Nacional Aut\'onoma de M\'exico ,\\
Apdo. Postal 20-364,  01000 M{\'e}xico D.F., \ M{\'e}xico.}}
\author{E. Peinado}
{address={Departamento de F\'{\i}sica Te\'orica,\\
Instituto de F\'{\i}sica, Universidad Nacional Aut\'onoma de M\'exico ,\\
Apdo. Postal 20-364,  01000 M{\'e}xico D.F., \ M{\'e}xico.}}

\begin{abstract}
\noindent
  We present an analysis of the theoretical neutrino mixing matrix, $V_{PMNS}^{th}$, previously derived in the framework of the minimal $S_3$-invariant extension of the Standard Model.
All entries in the neutrino mixing matrix, $V_{PMNS}^{th}$, the mixing
angles and the Majorana phases are given as exact, explicit analytical
functions of the mass ratios of the charged leptons and neutrinos, and
one Dirac phase, in excellent agreement with the the latest
experimental data. Here, it will be shown that all entries in
$V_{PMNS}^{th}$ are numerically very close to the tri-bimaximal form
of the neutrino mixing matrix, so that $V_{PMNS}^{th}$ may be written
as $V^{tri}+\Delta V_{PMNS}^{tri}$. The small correction $\Delta
V_{PMNS}^{tri}$ is expressed as a sum of two terms: first, a small
correction term proportional to $m_{e}/m_{\mu}$ depending only on the
charged lepton mass ratios and, second, a Cabbibo-like, small term, $\delta t_{12}$, which is a function of both the charged lepton and the neutrino mass ratios. 
\end{abstract}
\maketitle
\section{Introduction}
In the Standard Model (SM) the neutrinos are chiral and therefore massless,
but in the past nine years the experiments and observations related to the neutrino
physics showed us that neutrinos oscillate between states of well
defined flavours. In this way it was established beyond reasonable
doubt that neutrinos have non-vanishing masses~\cite{giunti}.
 
The masses of the neutrinos are at least five orders of magnitude smaller than the
electron mass, which is the lightest particle of all the fermions other
than the neutrinos. The small magnitude of neutrino masses is
naturally explained by assuming that the physical neutrinos are
Majorana fermions which acquire their masses via the see-saw mechanism~\cite{giunti,Langacker:2005pfa}.

The Standard Model explains almost all the experimental data at
laboratory energies but it cannot give mass to the neutrinos and,
even without the neutrino masses, it already has a large number of free
parameters. In order to accommodate the massive neutrinos and at the
same time reduce the number of free parameters in the theory, we
formulated a minimal $S_{3}$-invariant extension of the Standard
Model~\cite{Kubo:2003iw}. The flavour symmetry group of the extended
theory is the group $S_3$ of permutations of three objects, assumed to
be unbroken at the Fermi scale.

The symmetry group $S_3$ is the smallest non-Abelian group and it has
only doublet and singlet irreducible representations. In order to
have the theory invariant under $S_{3}$, we had to extend the
concept of flavour and family to the Higgs sector, that is, in the
extended form of the theory we have three Higgs $SU(2)_L$ doublets
that belong to one singlet and the two components of the one doublet
of $S_{3}$.

In this way, all the matter fields in the minimal $S_{3}$-invariant extension
of the Standard Model, that is, quarks, leptons and Higgs fields, are
in a reducible representation ${\bf 1_s \oplus 2}$ of $S_{3}$. We
constructed the most general Yukawa Lagrangian and the Higgs potential
invariant under this symmetry and obtained a generic form for the mass
matrices of the Dirac fermions~\cite{Kubo:2003iw}.

A further reduction of the number of parameters in the leptonic sector
may be achieved by means of an Abelian $Z_{2}$ symmetry. A set
of charge assignments of $Z_{2}$, compatible with the experimental
data on masses and mixings in the leptonic
sector is given in~\cite{Kubo:2003iw}. This allowed us to reparametrize the mass matrices of the charged leptons and neutrinos,
previously derived in~\cite{Kubo:2003iw}, in terms of their
eigenvalues and only one free parameter, the Dirac phase $\delta$. Then, we computed the neutrino mixing matrix, $V_{PMNS}$, and the neutrino
mixing angles and Majorana phases as functions of the masses of
charged leptons and neutrinos and only two undetermined phases. The numerical values of the reactor, $\theta_{13}$, and atmospheric,
$\theta_{23}$, mixing angles are determined only by the masses of the
charged leptons in very good agreement with experiment. The solar
mixing angle, $\theta_{12}$, is almost insensitive to the values of
the masses of the charged leptons, but its experimental value allowed
us to fix the scale and origin of the neutrino mass
spectrum~\cite{Felix:2006pn}.

More recently, we found exact, analytical expressions for the matrices of the
Yukawa couplings in the leptonic sector expressed as functions of the masses of charged
leptons and the vacuum expectation values of the Higgs bosons. With the help of the Yukawa matrices
we computed the branching ratios of some selected FCNC
processes~\cite{Mondragon:2007af}, and more recently, the contribution of the exchange of neutral flavour changing
scalars to the anomaly of the muon's magnetic moment~\cite{Mondragon:2007nk}. We found that the interplay of the $S_3\times Z_2$ flavour symmetry and the strong mass hierarchy of
charged leptons strongly suppress the FCNC processes in the leptonic
sector, well below the experimental upper bounds by many orders of
magnitude. The contribution of the FCNC to the anomaly, $a_{\mu}$, is at
most $6\%$ of the discrepancy between the experimental value and the
Standard Model prediction for $a_{\mu}$, which is a small but
non-negligible contribution~\cite{Mondragon:2007nk}.

In this short communication, we will show that all entries in
$V_{PMNS}^{th}$ derived in~\cite{Felix:2006pn} are numerically very close to the tri-bimaximal form
of the neutrino mixing matrix, so that, $V_{PMNS}^{th}$ may be written
as $V^{tri}+\Delta V_{PMNS}^{tri}$. The small correction $\Delta
V_{PMNS}^{tri}$ is expressed as a sum of two terms: first, a small
correction term proportional to $m_{e}/m_{\mu}$ depending only on the
charged lepton mass ratios and, second, a Cabbibo-like, small term,
$\delta t_{12}$, which is a function of both the charged lepton and
the neutrino mass ratios.

\section{Neutrino masses and mixings}

The neutrino mixing matrix $V_{PMNS}$ is the product
$U_{eL}^{\dagger}U_{\nu}K$, where $K$ is the diagonal matrix of the
Majorana phase factors, defined by
\be
diag(m_{\nu_{1}},m_{\nu_{2}},m_{\nu_{3}})=K^{\dagger}diag(|m_{\nu_{1}}|,|m_{\nu_{2}}|,|m_{\nu_{3}}|)K^{\dagger}.
\ee
Except for an overall phase factor, $e^{i\phi_{1}}$, which can be
ignored, $K$ is 
\be
K=diag(1,e^{i\alpha},e^{i\beta}).
\ee
The unitary matrix which diagonalizes the charged lepton mass matrix
is of the form ${\bf U}_{eL}=\Phi_{e} {\bf O}_{eL}$, where $\Phi_{e}$
is the diagonal phase matrix, $\Phi_{e}=diag(1,1,e^{i\delta_{e}})$, and
${\bf O}_{eL}$ is the orthogonal matrix
\begin{eqnarray}
{\bf O}_{eL}\approx
\left(
\ba{ccc}
\frac{1}{\sqrt{2}}x
\frac{(
1+2\tilde{m}_{\mu}^2+4x^2+\tilde{m}_{\mu}^4+2\tilde{m}_{e}^2
)}{\sqrt{1+\tilde{m}_{\mu}^2+5x^2-\tilde{m}_{\mu}^4-\tilde{m}_{\mu}^6+\tilde{m}_{e}^2+12x^4}}&
-\frac{1}{\sqrt{2}}\frac{(1-2\tilde{m}_{\mu}^2+\tilde{m}_{\mu}^4-2\tilde{m}_{e}^2)}{\sqrt{1-4\tilde{m}_{\mu}^2+x^2+6\tilde{m}_{\mu}^4-4\tilde{m}_{\mu}^6-5\tilde{m}_{e}^2}}
& \frac{1}{\sqrt{2}} \\ \\
-\frac{1}{\sqrt{2}}x
\frac{(
1+4x^2-\tilde{m}_{\mu}^4-2\tilde{m}_{e}^2
)}{\sqrt{1+\tilde{m}_{\mu}^2+5x^2-\tilde{m}_{\mu}^4-\tilde{m}_{\mu}^6+\tilde{m}_{e}^2+12x^4}}
&
\frac{1}{\sqrt{2}}\frac{(1-2\tilde{m}_{\mu}^2+\tilde{m}_{\mu}^4)}{\sqrt{1-4\tilde{m}_{\mu}^2+x^2+6\tilde{m}_{\mu}^4-4\tilde{m}_{\mu}^6-5\tilde{m}_{e}^2}}
& \frac{1}{\sqrt{2}} \\ \\
-\frac{\sqrt{1+2x^2-\tilde{m}_{\mu}^2-\tilde{m}_{e}^2}(1+\tilde{m}_{\mu}^2+x^2-2\tilde{m}_{e}^2)}{\sqrt{1+\tilde{m}_{\mu}^2+5x^2-\tilde{m}_{\mu}^4-\tilde{m}_{\mu}^6+\tilde{m}_{e}^2+12x^4}} & -x\frac{(1+x^2-\tilde{m}_{\mu}^2-2\tilde{m}_{e}^2)\sqrt{1+2x^2-\tilde{m}_{\mu}^2-\tilde{m}_{e}^2}}{\sqrt{1-4\tilde{m}_{\mu}^2+x^2+6\tilde{m}_{\mu}^4-4\tilde{m}_{\mu}^6-5\tilde{m}_{e}^2}} &\frac{\sqrt{1+x^2}\tilde{m}_{e}\tilde{m}_{\mu}}{\sqrt{1+x^2-\tilde{m}_{\mu}^2}}
\ea 
\right).
\label{unitary-leptons-2}
\end{eqnarray}
The Majorana neutrino mass matrix is obtained via the see-saw
mechanism ${\bf M_{\nu}} = {\bf M_{\nu_D}}\tilde{{\bf M}_R}^{-1} 
({\bf M_{\nu_D}})^T$, this matrix is diagonalized by a unitary matrix
\be
U_{\nu}^{T}M_{\nu}U_{\nu}=\mbox{diag}\left(|m_{\nu_{1}}|e^{i\phi_{1}},|m_{\nu_{2}}|e^{i\phi_{2}},|m_{\nu_{3}}|e^{i\phi_{\nu}}\right),
\label{diagneutrino}
\ee
where 
\be
U_{\nu}=\left(\ba{ccc} 
1& 0 & 0 \\
0 & 1 & 0 \\
0 & 0 & e^{i\delta_{\nu}} 
\ea\right)\left(
\begin{array}{ccc}
\cos \eta & \sin \eta & 0 \\
0 & 0 & 1 \\
-\sin \eta  & \cos \eta &0
\end{array}
\right),
\label{ununew}
\ee
and
\be
\ba{lr}
\sin^2\eta=\frac{m_{\nu_{3}}-m_{\nu_{1}}}{m_{\nu_{2}}-m_{\nu_{1}}},
&
\cos^2\eta=\frac{m_{\nu_{2}}-m_{\nu_{3}}}{m_{\nu_{2}}-m_{\nu_{1}}}.
\ea
\label{cosysin}
\ee 
The resulting theoretical mixing matrix, $V_{PMNS}^{th}=U_{eL}^{\dagger}U_{\nu}K$, is given by
\be
\hspace{-2.5cm}
V_{PMNS}^{th}=
\left(
\ba{ccc}
O_{11}\cos \eta + O_{31}\sin \eta e^{i\delta} & O_{11}\sin
\eta-O_{31} \cos \eta e^{i\delta} & -O_{21}  \\ \\
-O_{12}\cos \eta + O_{32}\sin \eta e^{i\delta} & -O_{12}\sin
\eta-O_{32}\cos \eta e^{i\delta} & O_{22} \\ \\
O_{13}\cos \eta - O_{33}\sin \eta e^{i\delta} & O_{13}\sin
\eta+O_{33}\cos \eta e^{i\delta} & O_{23} 
\ea
\right)
 \times K.
\label{vpmns}
\ee
To find the relation of our results
with the neutrino mixing angles, we make use of the equality of the
absolute values of the elements of $V_{PMNS}^{th}$ and
$V_{PMNS}^{PDG}$~\cite{PDG}, that is
\be
|V_{PMNS}^{th}|=|V_{PMNS}^{PDG}|.
\ee
This relation allowed us to derive expressions for the mixing angles
in terms of the charged lepton and neutrino masses~\cite{Felix:2006pn,Mondragon:2007af}.

The magnitudes of the reactor and atmospheric mixing angles,
$\theta_{13}$ and $\theta_{23}$, are determined by the masses of the
charged leptons only. Keeping only terms of order $(m_{e}^2/m_{\mu}^2)$ and
$(m_{\mu}/m_{\tau})^4$, we get
\be
\ba{lcr}
\sin \theta_{13}\approx \frac{1}{\sqrt{2}}x
\frac{(
1+4x^2-\tilde{m}_{\mu}^4)}{\sqrt{1+\tilde{m}_{\mu}^2+5x^2-\tilde{m}_{\mu}^4}}
 & ~\mbox{and}~&
\sin \theta_{23}\approx  \frac{1}{\sqrt{2}}\frac{1+\frac{1}{4}x^2-2\tilde{m}_{\mu}^2+\tilde{m}_{\mu}^4}{\sqrt{1-4\tilde{m}_{\mu}^2+x^2+6\tilde{m}_{\mu}^4}}.
\ea
\ee
The magnitude of the solar angle depends on the charged lepton and
neutrino masses, as well as the Dirac and Majorana phases
\be
\hspace{-2.0cm}
 |\tan \theta_{12}|^2= \frac{\displaystyle{m_{\nu_{2}}-m_{\nu_{3}}}}{
\displaystyle{m_{\nu_{3}}-m_{\nu_{1}}}}\left(\frac{1-2\frac{O_{11}}{O_{31}}\cos \delta\sqrt{\frac{\displaystyle{m_{\nu_{3}}-m_{\nu_{1}}}}{
\displaystyle{m_{\nu_{2}}-m_{\nu_{3}}}}}+\left(\frac{O_{11}}{O_{31}}\right)^2\frac{\displaystyle{m_{\nu_{3}}-m_{\nu_{1}}}}{
\displaystyle{m_{\nu_{2}}-m_{\nu_{3}}}}}{1+2\frac{O_{11}}{O_{31}}\cos \delta\sqrt{\frac{\displaystyle{m_{\nu_{2}}-m_{\nu_{3}}}}{
\displaystyle{m_{\nu_{3}}-m_{\nu_{1}}}}}+\left(\frac{O_{11}}{O_{31}}\right)^2\frac{\displaystyle{m_{\nu_{2}}-m_{\nu_{3}}}}{
\displaystyle{m_{\nu_{3}}-m_{\nu_{1}}}}
}\right)
.
\label{tan2}
\ee
From this expression, we found $m_{\nu_3}$ as function of the
solar angle, the squared neutrino mass differences and $\phi_{\nu}$,
and, thus, fixed the origin and scale of the neutrino masses.

We are interested in the relation between our mixing matrix in
Eq. (\ref{vpmns}) and the tri-bimaximal
pattern~\cite{Harrison:2002er}. In order to obtain this relation, it
would be convenient to write $\tan \theta_{12}$ as
\be
\tan \theta_{12}=\frac{1}{\sqrt{2}} +\delta t_{12},
\label{theta}
\ee
where $\delta t_{12}$ is a small quantity of the order of $6\%$ of the
tri-bimaximal value.

\section{Deviation  of the mixing matrix ${\bf V}_{PMNS}^{th}$ from the tri-bimaximal form}

The previous results on neutrino masses and mixings weakly depend on the Dirac
phase $\delta$, for simplicity we will assume in this work that
$\delta=\pi/2$. From the expression (\ref{vpmns}), and the
Eq. (\ref{theta}), we may write the mixing matrix as follows,
\be
{\bf V}_{PMNS}^{th}={\bf V}_{PMNS}^{tri}+
  \Delta V_{PMNS}^{tri},
\ee
where the tri-bimaximal form ${\bf
  V}_{PMNS}^{tri}$~\cite{Harrison:2002er} is
\be
{\bf V}_{PMNS}^{tri}=\left(\ba{ccc}
\sqrt{\frac{2}{3}} &\sqrt{\frac{1}{3}} & 0\\
-\sqrt{\frac{1}{6}} &\sqrt{\frac{1}{3}} & -\sqrt{\frac{1}{2}}\\
-\sqrt{\frac{1}{6}} &\sqrt{\frac{1}{3}} & \sqrt{\frac{1}{2}}
\ea\right),
\label{tribimax}
\ee
and 
\be\Delta{\bf  V}_{PMNS}^{tri}=\Delta {\bf  V}_{e}+ \delta t_{12}\frac{(\sqrt{2}+\delta t_{12})}{g(\delta t_{12})}\Delta{\bf
  V}_{\nu},
\label{del}
\ee
where 
\be
g(\delta t_{12})=1+\frac{2}{3}\delta t_{12}(\sqrt{2}+\delta t_{12}).
\ee 
Comparing the expression (\ref{vpmns}) for the neutrino mixing matrix,
with the tri-bimaximal form (\ref{tribimax}) we get,
\be
\Delta {\bf V}_{e}\approx\left(\ba{ccc}
-\frac{2}{3}\frac{{s_{13}^2}}{1+c_{13}} &
-\frac{1}{3}\frac{{s_{13}^2}}{1+c_{13}} &s_{13} \\
\frac{5}{2\sqrt{6}}\frac{{x^2}}{1+\sqrt{1+\frac{5}{2}x^2}} & \frac{1}{4}
\sqrt{\frac{1}{3}}\frac{{x^2}}{1+\sqrt{1+\frac{1}{4}x^2}} &
-\frac{1}{2\sqrt{2}}\frac{{x^2}}{\sqrt{1-4\tilde{m}_{\mu}^2+x^2}}
\\ 
\sqrt{\frac{1}{6}}\frac{x^2}{1+\sqrt{1+x^2}} &
\frac{1}{4}\sqrt{\frac{1}{3}}\frac{x^2}{1+\sqrt{1+\frac{1}{4}x^2}} & 0
\ea\right),
\label{deltae}
\ee
where $x=m_{e}/m_{\mu}$, $\tilde{m}_{\mu}=m_{\mu}/m_{\tau}$ and $s_{13}\approx 1/\sqrt{2}x (
1+4x^2-\tilde{m}_{\mu}^4)/\sqrt{1+\tilde{m}_{\mu}^2+5x^2-\tilde{m}_{\mu}^4}$.

Notice that all entries in Eq. (\ref{deltae}) are proportional to
$x^2$ except for the $(\Delta {\bf V}_{e})_{13}$ which is proportional to
$x$. Therefore, in the limit of vanishing electron mass, $\Delta{\bf
  V}_{e}\rightarrow 0 $. 

The matrix $\Delta{\bf
  V}_{\nu}$ can be written as
\be
\Delta{\bf V}_{\nu}= 
\left(\ba{ccc}
-
\left(\frac{2}{3}\right)^{3/2}\frac{1-\frac{s_{13}^2}{1+c_{13}}}{1+\sqrt{1-\frac{2}{3}\frac{\delta t_{12}(\sqrt{2}+\delta t_{12})}{g(\delta t_{12})}}}&
\left(\frac{1}{3}\right)^{3/2}\frac{1-\frac{s_{13}^2}{1+c_{13}}}{1+\sqrt{1+\frac{1}{3}\frac{\delta t_{12}(\sqrt{2}+\delta t_{12})}{g(\delta t_{12})}}}&0\\
\left(\frac{2}{3}\right)^{3/2}\frac{1-2x^2}{\sqrt{1+\frac{5}{2}x^2}\left(1+\sqrt{1+\frac{4}{3}\frac{\delta t_{12}(\sqrt{2}+\delta t_{12})}{g(\delta t_{12})}\frac{1-2x^2}{1+\frac{5}{2}x^2}}\right)}&
-\frac{2}{3\sqrt{3}}\frac{1-2x^2}{\sqrt{1+\frac{1}{4}x^2}\left(1+\sqrt{1-\frac{2}{3}\frac{\delta t_{12}(\sqrt{2}+\delta t_{12})}{g(\delta t_{12})}\frac{1-2x^2}{1+\frac{1}{4}x^2}}\right)}&
0\\
\left(\frac{2}{3}\right)^{3/2}
\frac{1+\frac{1}{4}x^2}{\sqrt{1+x^2}\left(1+\sqrt{1+\frac{4}{3}\frac{\delta t_{12}(\sqrt{2}+\delta t_{12})}{g(\delta t_{12})}\frac{1+\frac{1}{4}x^2}{1+x^2}}\right)}&
-\frac{2}{3\sqrt{3}}\frac{1+x^2}{\sqrt{1+\frac{1}{4}x^2}\left(1+\sqrt{1+\frac{2}{3}\frac{\delta t_{12}(\sqrt{2}+\delta t_{12})}{g(\delta t_{12})}\frac{1+x^2}{1+\frac{1}{4}x^2}}\right)}&0\ea \right)
\label{delnu}
\ee
notice that all the entries in the third column of $\Delta{\bf
  V}_{\nu}$ vanish. 

From Eq. (\ref{del}) and Eq. (\ref{deltae}) we can see that in the
limit of $m_e=0$ and $\delta t_{12}=0$ the deviation from the
tri-bimaximal pattern is exactly zero, that is
\be
\Delta{\bf  V}_{PMNS}^{tri}=0.
\ee

The value for $\delta t_{12}$ fixes the scale and the origin
of the neutrino masses. If we take for $\delta t_{12}$ the
experimental central value $\delta t_{12}\approx -0.04$, we
obtain~\cite{Mondragon:2007af}
\[
\ba{l}
|m_{\nu_{2}}|\approx0.056eV\\ |m_{\nu_{1}}|\approx 0.055eV,
\ea
\]
and
\[
|m_{\nu_{3}}|\approx 0.022eV.
\]

When we take for $\delta t_{12}$ the tri-bimaximal value $\delta t_{12}=0$, the neutrino masses are
\be
\ba{lccr}
m_{\nu_1}= 0.0521~eV &  m_{\nu_2}=0.0528 ~eV &~\mbox{and}~& m_{\nu_3}= 0.0178~eV\ea
\ee
In both cases the $S_{3}$ invariant extension of the SM predicts an
inverted hierarchy. Since the tri-bimaximal value of $\delta t_{12}$
differs from the experimental central value by less than $6\%$ of
$\tan \theta_{12}$, the difference in the corresponding numerical 
values of the neutrino masses are not significative within
the present experimental uncertainties.

\section{Conclusions}

In the minimal $S_{3}$-invariant extension of the SM, the flavour symmetry group $S_{3}
\times Z_{2}$ relates the mass spectrum and mixings, and reduces the
number of free parameters in the leptonic sector of the theory. This allowed us to
predict two mixing angles, $\theta_{13}$ and $\theta_{23}$, as
function of the charged lepton masses only, in excellent agreement with the
experimental values, while the other neutrino mixing angle,
$\theta_{12}$,
depends also on the neutrino mass spectrum. The tangent of the solar
angle,$\tan \theta_{12}$, fixes the scale and origin of the neutrino
masses which has an inverted hierarchy. In this model, we found that
the deviation of the theoretical neutrino mixing matrix,$V_{PMNS}^{th}$,
from the tri-bimaximal pattern is very small. In this form of the
theory, the flavour violating processes are strongly suppressed by the
flavour symmetry $S_{3}\times Z_{2}$ and the strong mass hierarchy of
the charged leptons. Processes that proceed through the exchange of
flavour changing neutral currents give information on the vacuum
expectation values for the scalar Higgs
bosons. The contribution of FCNC to the magnetic moment anomaly of the
muon is at most $6\%$ of the discrepancy between the experimental value and the
Standard Model prediction, this contribution is small but non-negligible~\cite{Mondragon:2007nk}.
\section{Acknowledgements}
This work was partially supported by CONACYT M\'exico under contract
No 51554-F and by DGAPA-UNAM under contract PAPIIT-IN115207-2.


\begin{thebibliography}{20}
\bibitem{giunti} See for instance C. Giunti, ``Neutrino Physics'' in these
  proceedings.
\bibitem{Langacker:2005pfa} 
  P.~Langacker, J.~Erler and E.~Peinado,
  ``Neutrino physics,''
  J.\ Phys.\ Conf.\ Ser.\  {\bf 18} (2005) 154; [arXiv:hep-ph/0506257].
\bibitem{Kubo:2003iw}
  J.~Kubo, A.~Mondragon, M.~Mondragon and E.~Rodriguez-Jauregui,
  Prog.\ Theor.\ Phys.\  {\bf 109} (2003) 795
  [Erratum-ibid.\  {\bf 114} (2005) 287]
  [arXiv:hep-ph/0302196].
\bibitem{Felix:2006pn}
  O.~Felix, A.~Mondragon, M.~Mondragon and E.~Peinado,
  ``Neutrino masses and mixings in a minimal S(3)-invariant extension of the
  standard model,''
  AIP Conf.\ Proc.\  {\bf 917} (2007) 383
  [Rev.\ Mex.\ Fis.\  {\bf S52N4} (2006) 67];
  [arXiv:hep-ph/0610061].

\bibitem{Mondragon:2007af}
  A.~Mondragon, M.~Mondragon and E.~Peinado,
  Phys.\ Rev.\  D {\bf 76} (2007) 076003
  [arXiv:0706.0354 [hep-ph]].

\bibitem{Mondragon:2007nk}
  A.~Mondragon, M.~Mondragon and E.~Peinado,
  ``$S_3$-flavour symmetry as realized in lepton flavour violating processes,''
  arXiv:0712.1799 [hep-ph].


\bibitem{Harrison:2002er}
  P.~F.~Harrison, D.~H.~Perkins and W.~G.~Scott,
  {\it Phys.\ Lett.}  B {\bf 530} (2002) 167
  arXiv:hep-ph/0202074.

\bibitem{PDG} W-M Yao {\it et al.} [Particle Data Group],{\it J. Phys. G: Nucl. Part. Phys.}
  {\bf 33} (2006) 1.
\end{thebibliography}
\end{document}